\documentclass[twocolumn,showpacs,%
nofootinbib,aps,superscriptaddress,%
eqsecnum,prd,notitlepage,showkeys,10pt]{revtex4-1}
\usepackage{amsmath,amssymb,amsfonts}
\usepackage{graphicx}
\usepackage{textcomp}
\usepackage{xcolor}

\usepackage[noend]{algpseudocode}
\usepackage{algorithm}
\usepackage{comment}
\usepackage[hidelinks]{hyperref}
\makeatletter % changes the catcode of @ to 11
\newcommand{\linebreakand}{%
\end{@IEEEauthorhalign}
\hfill\mbox{}\par
\mbox{}\hfill\begin{@IEEEauthorhalign}
}
\makeatother % changes the catcode of @ back to 12

\def\BibTeX{{\rm B\kern-.05em{\sc i\kern-.025em b}\kern-.08em
		T\kern-.1667em\lower.7ex\hbox{E}\kern-.125emX}}
\begin{document}
	
	\title{
		A Control Flow based Static Analysis of GRAFCET using Abstract Interpretation
	}

%	\author{
%		\IEEEauthorblockN{
%			Aron Schnakenbeck\textsuperscript{1}, 
%			Robin Mroß\textsuperscript{2}, 
%			Marcus Völker\textsuperscript{2}, 
%			Stefan Kowalewski\textsuperscript{2}, 
%			Alexander Fay\textsuperscript{1}
%		}
%		\IEEEauthorblockA{
%			\textsuperscript{1}Institute of Automation, Helmut Schmidt University, Hamburg, Germany\\
%			{\{aron.schnakenbeck, alexander.fay\}@hsu-hh.de}\\
%			\textsuperscript{2} Computer Science 11 - Embedded Software, RWTH Aachen University, Aachen, Germany\\
%			{\{mross, voelker, kowalewski\}@embedded.rwth-aachen.de}
%		}	
%	}

\author{Aron Schnakenbeck}
\affiliation{
			\textit{Institut für Automatisierungstechnik} \\
			\textit{Helmut-Schmidt-Universität}\\
			22043 Hamburg, Germany \\
			\{aron.schnakenbeck, alexander.fay\}@hsu-hh.de
			%		https://orcid.org/0000-0002-3507-2851
		}
\author{Robin Mroß}
\affiliation{
			\textit{Lehrstuhl Informatik 11} \\
			\textit{RWTH Aachen University}\\
			52074 Aachen, Germany \\
			\{mross, voelker, kowalewski\}@embedded.rwth-aachen.de
	%%		https://orcid.org/0000-0002-1007-5597
}
\author{Marcus Völker}
\affiliation{
	\textit{Lehrstuhl Informatik 11} \\
	\textit{RWTH Aachen University}\\
	52074 Aachen, Germany \\
	\{mross, voelker, kowalewski\}@embedded.rwth-aachen.de
	%%		https://orcid.org/0000-0002-1007-5597
}
\author{Stefan Kowalewski}
\affiliation{
	\textit{Lehrstuhl Informatik 11} \\
	\textit{RWTH Aachen University}\\
	52074 Aachen, Germany \\
	\{mross, voelker, kowalewski\}@embedded.rwth-aachen.de
	%%		https://orcid.org/0000-0002-1007-5597
}
\author{Alexander Fay}
\affiliation{
	\textit{Institut für Automatisierungstechnik} \\
	\textit{Helmut-Schmidt-Universität}\\
	22043 Hamburg, Germany \\
	\{aron.schnakenbeck, alexander.fay\}@hsu-hh.de
	%		https://orcid.org/0000-0002-3507-2851
}

%\author{
%		\IEEEauthorblockN{Aron Schnakenbeck}
%	\IEEEauthorblockA{
%		\textit{Institut für Automatisierungstechnik} \\
%		\textit{Helmut-Schmidt-Universität}\\
%		22043 Hamburg, Germany \\
%		aron.schnakenbeck@hsu-hh.de
%		%		https://orcid.org/0000-0002-3507-2851
%	}
%	\and
%	\IEEEauthorblockN{Robin Mro{\ss}}
%	\IEEEauthorblockA{
%		\textit{Lehrstuhl Informatik 11} \\
%		\textit{RWTH Aachen University}\\
%		52074 Aachen, Germany \\
%%		https://orcid.org/0000-0002-1007-5597
%	}
%	\and
%	\IEEEauthorblockN{Marcus V\"{o}lker}
%	\IEEEauthorblockA{
%		\textit{Lehrstuhl Informatik 11} \\
%		\textit{RWTH Aachen University}\\
%		52074 Aachen, Germany \\
%%		https://orcid.org/0000-0001-7348-0146
%	}
%	\and
%\IEEEauthorblockN{Stefan Kowalewski}
%\IEEEauthorblockA{
%	\textit{Lehrstuhl Informatik 11} \\
%	\textit{RWTH Aachen University}\\
%	52074 Aachen, Germany \\
%	%		https://orcid.org/0000-0002-7184-4804
%}
%	\and
%	\linebreakand
%	\IEEEauthorblockN{Alexander Fay}
%	\IEEEauthorblockA{
%		\textit{Institut für Automatisierungstechnik} \\
%		\textit{Helmut-Schmidt-Universität}\\
%		22043 Hamburg, Germany \\
%%		https://orcid.org/0000-0002-1922-654X
%	}
%
%}

	\begin{abstract}
		The graphical modeling language GRAFCET is used as a formal specification language in industrial control design. This paper proposes a static analysis approach based on the control flow of GRAFCET using abstract interpretation to allow verification on specification level. GRAFCET has different elements leading to concurrent behavior, which in general results in a large state space. 
		To get precise results and reduce the state space, we propose an analysis suitable for GRAFCET instances without concurrent behavior. We point out how to check for the absence of concurrency and present a flow-sensitive analysis for these GRAFCET instances.		
		The proposed approach is evaluated on an industrial-sized example.
	\end{abstract}

		\maketitle
%	\begin{IEEEkeywords}
%		Static Analysis, GRAFCET, Abstract Interpretation, Industrial Automation
%	\end{IEEEkeywords}
%	
	
	\section{Introduction}

In industrial automation, Programmable Logic Controllers (PLC) are widely used. To design the control code running on a PLC a beneficial approach is to use formal means in order to first specify the logical behavior of the PLC before implementing the control code. 
Using a formal specification in the design phase has multiple advantages like using the specification as documentation and communication tool, allowing an automatic transformation into control code and applying formal verification at specification level. 
One such means is GRAFCET according to IEC 60848 \cite{iec60848} a graphical, semi-formal, domain-specific language to model control code of PLCs.
As we have shown in \cite{Mross.22}, GRAFCET is used in several industrial domains and is widely known in the respective areas. This acceptance of GRAFCET might improve the acceptance of formal methods in the respective domains, which is still a problem \cite{VogelHeuser.14}.
Although GRAFCET adapts concepts of Petri nets -~like transitions and steps, connected alternately by arcs~- it provides a considerable number of additional modeling mechanisms like hierarchical structuring of the specification which allow for compact modeling of complex systems \cite{Mross.22}. 

Regarding the application of formal methods to GRAFCET specifications, there is preliminary work by Julius et al. \cite{Julius.17} to allow a code generation of such hierarchical GRAFCET specifications to PLC-code.
Because the work presented by Julius et al. does not cover verification of the Grafcets (the term \textit{Grafcet} refers to an instance of GRAFCET), we extend the approach by a formal verification of GRAFCET.
A verification on specification level has the advantage of finding possible design errors early in the design process, given that the costs of correcting errors in software systems increases exponentially as the development phase progresses \cite{Boehm.1981}.

The verification approach proposed in this work is a static analysis using abstract interpretation based on the control flow of GRAFCET.
We will compare the proposed approach to other possible approaches in Section~\ref{sec:relWork}, followed by the preliminaries on GRAFCET and abstract interpretation in Section~\ref{sec:prelim}. The behavior of a Grafcet depends on its state which is composed of the possible active steps %(the situation of the Grafcet)
and the possible assignments of the internal variables. 
In Section~\ref{ssec:problem} we will point out how these states of a Grafcet can be approximated by its control flow which is only possible by ensuring the absence of concurrency. We will present what elements of the GRAFCET standard result in concurrent behavior and how to ensure its absence. Section~\ref{ssec:flowsens} presents the actual analysis.
We end with evaluating the proposed analysis on a practical example and showing properties to be analyzed (Section \ref{sec:eval}) before giving a conclusion (Section \ref{sec:conclusion}).
	\section{Related work}
\label{sec:relWork}

In the domain of industrial automation an important requirement for the application of formal means is that it is appropriate for craftsman and that it needs to be performed under time pressure \cite{VogelHeuser.14}. Therefore an automatic verification approach that does not require an expert is beneficial in contrast to approaches like theorem proving. 

For verifying GRAFCET there are approaches suitable for model checking, such as translating hierarchical Grafcets into time Petri nets by Sogbohossou et al. \cite{Sogbohossou.20} and recently transforming Grafcets into Guarded Action Language (GAL) resulting in a transition system by Mroß et al. \cite{Mross.22}. 
Utilizing a model checking approach allows for an exhaustive exploration of the model but has the downside of resulting in a state space explosion.

Very few approaches are presented for analyzing GRAFCET without applying model checking.
A structural analysis regarding the hierarchical dependencies between modules of the Grafcets (called \textit{partial Grafcets}) has been presented by Lesage et al. \cite{Lesage.93}. The authors provide an analysis to ensure that the hierarchical dependencies form a partial order. Moreover, Lesage et al. \cite{Lesage.96} provide an analysis of the GRAFCET-specific expressions by extending the Boolean algebra by events represented by rising and falling edges of Boolean signals in GRAFCET. This allows the user to check syntactic properties of transition conditions.  
None of the presented approaches allows for a static analysis of the run-time behavior.  

%%%%%Abstrakte Interpertation
%Arbeiten die genutz werden
A different approach and the main idea pursued in this work is to approximate the state space of the Grafcets by means of abstract interpretation proposed by Cousot et al. \cite{Cousot.77}. 
%Simon
An application of abstract interpretation to Sequential Function Chart (SFC), a graphical programming language in the field of industrial automation, has been proposed by Simon et al. \cite{Simon.16}. The analysis builds the reachability graph of the SFCs and calculates abstract values of the variables as pre- and post-conditions of the reachable states.
%Peleska
An analysis of statecharts, a specification language comparable to GRAFCET, with automatically generated test cases, has been proposed by Peleska et al. \cite{Peleska.11}. The proposed algorithm calculates an abstract computation sequence of the statechart using abstract interpretation. 
Both approaches presented in \cite{Simon.16, Peleska.11} are based on building some kind of reachability graph over the set of steps (for state diagrams called \textit{control states}) and approximating the internal variables using abstract interpretation. This might result in a huge state space depending on how many parallel steps are present in the system to be modeled.

To achieve the most scalable results, also in comparison to model checking approaches like \cite{Mross.22}, we propose a static analysis using abstract interpretation based on the control flow of GRAFCET.
A standard algorithm applying abstract interpretation to the control flow of a sequential program has been described in \cite{Nielson.99} and shown in the next section. 
%mulit-threading
The algorithm was extended to apply it to concurrent multi-threading programs e.g., by Kusano et al. \cite{Kusano.16}. Analyzing multi-threading programs results in an interplay of multiple Control Flow Graphs (CFG). Although this might be a promising approach for notably partial Grafcets, there are structures in Grafcet that are not comparable with a sequential CFG, which are shown in Section \ref{ssec:problem}.

	\section{Preliminaries}
\label{sec:prelim}

\begin{figure}[t]
	\centering
	\includegraphics[width=.4\columnwidth]{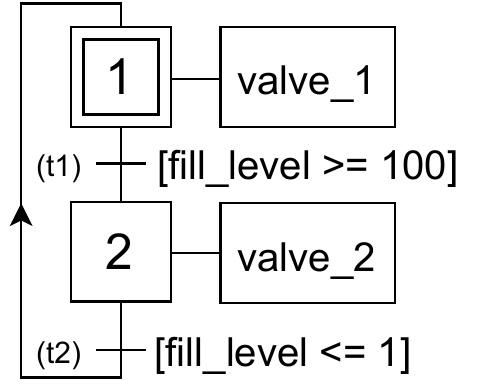}
	\caption{Illustrative example of a Grafcet.}
	\label{fig:exampleGrafcet}
\end{figure}

The goal of this work is to adapt abstract interpretation based on a CFG representing a program to GRAFCET as formalized in Section \ref{ssec:syntax}. To  explain the analogies as well as differences between a CFG and GRAFCET in Section \ref{sec:contribution} we start by providing the preliminaries on abstract interpretation based on a CFG in Section \ref{ssec:abstractInterpretation}.

\subsection{Syntax of IEC 60848 GRAFCET}
\label{ssec:syntax}
Since the GRAFCET standard does not define the syntax and semantics of GRAFCET sufficiently for formal verification, we use in this work the formalization proposed by Mroß et al. \cite{Mross.22} to explain the concepts of GRAFCET that are important for this contribution.

A Grafcet $G = (V_\mathit{in}, V_\mathit{int}, V_\mathit{out}, C)$ comprises a set of partial Grafcets $C \neq \emptyset$ with globally available sets of input variables $V_\mathit{in}$, internal variables $V_{int}$ and output variables $V_\mathit{out}$. Variables can either be Boolean or integral, i.e., $v$ is assigned a value of $\mathbb{Z}$ for all $v \in V_\mathit{in} \cup V_\mathit{int} \cup V_\mathit{out}$ with Boolean variables being limited to the set $\{0, 1\}$. 
%We denote by $V_{inB} \subseteq V_{in}$ ($V_{intB} \subseteq V_{int}$, $V_{outB} \subseteq V_{out}$) the set of Boolean input (internal, output) variables. 
Given these variables, we can construct Boolean expressions with usual relational symbols (such as $=$ and $\le$) and Boolean operators (such as disjunction $\lor$ and negation $\lnot$). A variable may change values caused by an event. 
By $\mathit{CND}$ we denote the set of all Boolean expressions over variables in $G$. 
Every partial Grafcet $c \in C$ is a 6-tuple $c = (S, I, E, M, T, A)$, where 
\begin{itemize}
	\item $S$ is a finite set of steps, each of which is either active or inactive,
	\item $I \subseteq S$ is the set of initial steps,
	\item $E \subseteq S \times C$ is the set of enclosing steps,
	\item $M \subseteq S$ is the set of marked steps,
	\item $T \subseteq \mathcal{P}(S) \times \mathcal{P}(S) \times \mathit{CND}$ is the set of transitions,
	\item $A$ is a set of actions.
\end{itemize}
Fig. \ref{fig:exampleGrafcet} shows an illustrative example of a partial Grafcet with two steps $S = \{1, 2\}$ one of which is an initial step $ I = \{1\}$ and two transitions $T = \{\mathrm{t1}, \mathrm{t2}\}$ 
%$T=\{\langle\{1\}, \{2\}, fill\_level \geq 100 \rangle,\langle\{2\}, \{1\}, fill\_level \leq 1\rangle\}$ 
as well as two actions associated to step 1 and step 2.

We use the notation $S_{c}$, $I_{c}$, $E_{c}$, $M_{c}$, $T_{c}$, $A_{c}$ to refer to the respective sets of a given partial Grafcet $c \in C$. The set $M_{c}$ describes the steps that are activated by the enclosing step. Every $e \in E_{c}$ describes an enclosing step, which translates formally to $e = (s, c_\mathit{enc})$ for a $s \in S_{c}$ and a partial Grafcet $c_\mathit{enc} \in C$. If an enclosing step becomes active, it activates all steps $m \in M_{c_\mathit{enc}}$. If an enclosing step becomes inactive, it deactivates all steps $s \in S_{c_\mathit{enc}}$. We say that $c$ is \textit{enclosed} iff $M_{c} \neq \emptyset$. 
%Further we have disjoint sets of steps, that is  $S_{c} \cap S_{c'} = \emptyset$ for every $c' \in C$ with $c \neq c'$. 
Every step $s \in S_{c}$ induces a new Boolean variable $x_{s}$ which indicates the activation status of $s$ and is true iff the step is active in the current situation. These variables can be used in Boolean expressions $\mathit{CND}$. 

A transition $t \in T_{c}$ is a triple $t = (\bullet t, t \bullet, b)$, where 
$\bullet t \subseteq S_{c}$ is the set of immediately preceding steps, 
$t \bullet \subseteq S_{c}$ is the set of immediately succeeding steps, 
$\bullet t \neq \emptyset \lor t \bullet \neq \emptyset$
and $b \in \mathit{CND}$ is the transition condition.
We also call $\bullet t$ the \textit{upstream} and $t \bullet$ the \textit{downstream} of $t$.
We say that $t$ is \textit{enabled} if $x_{s}$ is true for every $s \in \bullet t$. We say that $t$ can \textit{fire} if it is enabled and $b$ is true.

Finally, we formalize the set of actions $A_{c}$. The standard defines different types of actions: continuous actions ($A_{cont}$), stored actions ($A_\mathit{stor}$) and forcing orders ($A_\mathit{fo}$). 
These sets are assumed to be disjoint. Let $A_{c} = A_\mathit{cont} \cup A_\mathit{stor} \cup A_{\mathit{fo}}$. Every element of $A_\mathit{cont}$ is a triple $(s, v, b)$, where 
$s \in S_{c}$ is the associated step, 
$v \in V_\mathit{out}$ is an output variable which must be Boolean and
$b \in \mathit{CND}$ is the action condition.
We say that a continuous action is \textit{active} if $x_{s}$ and $b$ are true. Several partial Grafcets in $G$ may employ continuous actions on the same output variable $v$. In this case, $v$ is set to true if at least one of these continuous actions is active. Note that $v$ can not be used by any stored action.
Every element of $A_\mathit{stor}$ is a tuple $(s, v, val, b)$, where 
$s \in S_{c}$ is the associated step, 
$v \in V_\mathit{int} \cup V_\mathit{out}$ is an internal or output variable, 
$val$ is an expression yielding a value in the respective domain, e.g., $val \in \mathbb{Z}$ and 
$b \in \mathit{CND}$ is the action condition.
A stored action sets $v$ to $val$ if $x_{s}$ and $b$ are true. This also allows to model actions on activation and deactivation of a step, as introduced by the standard. 
Finally, every element of $A_\mathit{fo}$ is a tuple $(s, c_\mathit{forced}, S)$, where 
$s \in S_{c}$ is the associated step, 
$c_\mathit{forced} \in C$ is the partial Grafcet which is to be forced and 
$S \in (\mathcal{P}(S_{c_\mathit{forced}}) \cup \{*, \mathit{init}\})$.
A forcing order is regarded as a special kind of continuous action. It is active while $x_{s}$ is true and forces $c_\mathit{forced}$ into the situation specified by $S$. If $S = *$, then the current situation in $c_\mathit{forced}$ is retained for as long as $s$ is active. If $S = \mathit{init}$ then $c_\mathit{forced}$ is set to its initial situation. Otherwise, it is set to the specified situation (element of the power set $\mathcal{P}(S_{c_\mathit{forced}})$). 
\begin{figure}[t]
   \centering
   \includegraphics[width=.9\columnwidth]{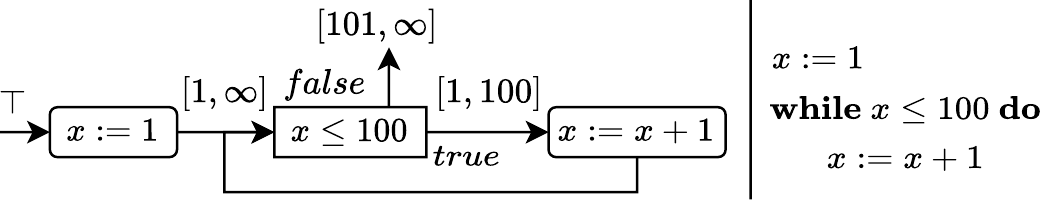}
   \caption{Example CFG and its corresponding program adapted from \cite{Cousot.77}.}
   \label{fig:cfg}
\end{figure}

%======================================================
%abstract interpretation
%======================================================
\subsection{Abstract Interpretation}
\label{ssec:abstractInterpretation}

In this section we provide the preliminaries on abstract interpretation based on a CFG. 
The nodes of such a CFG represent the instructions of the represented program and the edges represent the control flow paths \cite{Allen.70}.
Fig. \ref{fig:cfg} shows an example CFG adapted from \cite{Cousot.77}. The program has three statements, two assignments and a condition, as well as an entry (exit) point denoted by the incoming (outgoing) arrow. Note that we restrict ourselves in this work to sequential programs, i.e., programs written for example in C running on a single thread.

Abstract interpretation allows to approximate concrete variable values by an element of an abstract domain. The abstract domain in this work is the interval lattice. For every program point, every variable has a lower and an upper bound, e.g., a variable $x=[1, 100]$. All possible intervals, meet ($\sqcap$) and join ($\sqcup$) operators, a partial ordering ($\sqsubseteq$) as well as a bottom ($\bot = \emptyset$) and a top ($\top = [-\infty, \infty]$) element form the lattice. 
To apply abstract interpretation based on a CFG the worklist algorithm in Alg. \ref{alg:aiCFG} \cite{Nielson.99} is a standard algorithm ($A \triangleleft a$ is short for $A \leftarrow A \cup a$). It calculates an abstract environment $\mathit{Env}(n)$ for every node $n$ of the CFG, just before $n$ is executed. As long as a node $n$ is on the worklist $W$ the algorithm calculates a new environment $e$ by executing $n$ in the abstract domain using the \textsc{Transfer} function. The \textsc{Transfer} function performs in our case interval arithmetic calculations for assignments (e.g., $[1, 100] + [1, 1] = [2, 101]$ continuing the example above and executing $x := x + 1$), or it uses the meet operator to intersect with the interval of the condition (e.g., $[2, 101] \sqcap [-\infty, 100] = [2, 100]$ continuing the example further and executing $x \leq 100$).
Unless the analysis has stabilized (i.e., $e \sqsubseteq \mathit{Env}(n')$) the algorithm joins the calculated environment with the environment of the successor nodes $n'$ (e.g., $[2, 101] \sqcup [1, 100] = [1, 101]$) and adds them to the worklist.
To speed up the calculations and guarantee termination in case of loops, a widening operator \cite{Cousot.77} can be used.

\begin{algorithm}[H]
	\caption{Abstract interpretation based on a CFG \cite{Nielson.99}}
	\label {alg:aiCFG}
	\begin{algorithmic}[1]
  \Function{AbstInt}{$\mathit{Cfg}$: CFG}
    \State $\mathit{Env}(n)$ is initialized to $\top$ if $n \in \textsc{Entry}(\mathit{Cfg})$, else to $\bot$
    \State $W \leftarrow \textsc{Entry}(\mathit{Cfg})$
    \While {$W \neq \emptyset$}%{$\exists n \in W$}
        %\State $W \leftarrow W \backslash \{n\}$
        \State $n \leftarrow W.poll$%   W \leftarrow W \backslash \{n\}$
        \State $e \leftarrow \textsc{Transfer}(n, \mathit{Env}(n))$
        \ForAll{$n' \in \textsc{Succs}(\mathit{Cfg},n)$ such that $e \not\sqsubseteq \mathit{Env}(n')$}
            \State $\mathit{Env}(n') \leftarrow \mathit{Env}(n') \sqcup e$
            %\State $W \leftarrow W \cup {n'}$
            \State $W \triangleleft {n'}$
        \EndFor     
    \EndWhile
    \State \textbf{return} $\mathit{Env}$
    \EndFunction
	\end{algorithmic}
\end{algorithm}

%-----------------------------------------
\section{Applying abstract interpretation to GRAFCET}
\label{sec:contribution}
\begin{figure*}[t]
	\centering
	\includegraphics[width=.8\textwidth]{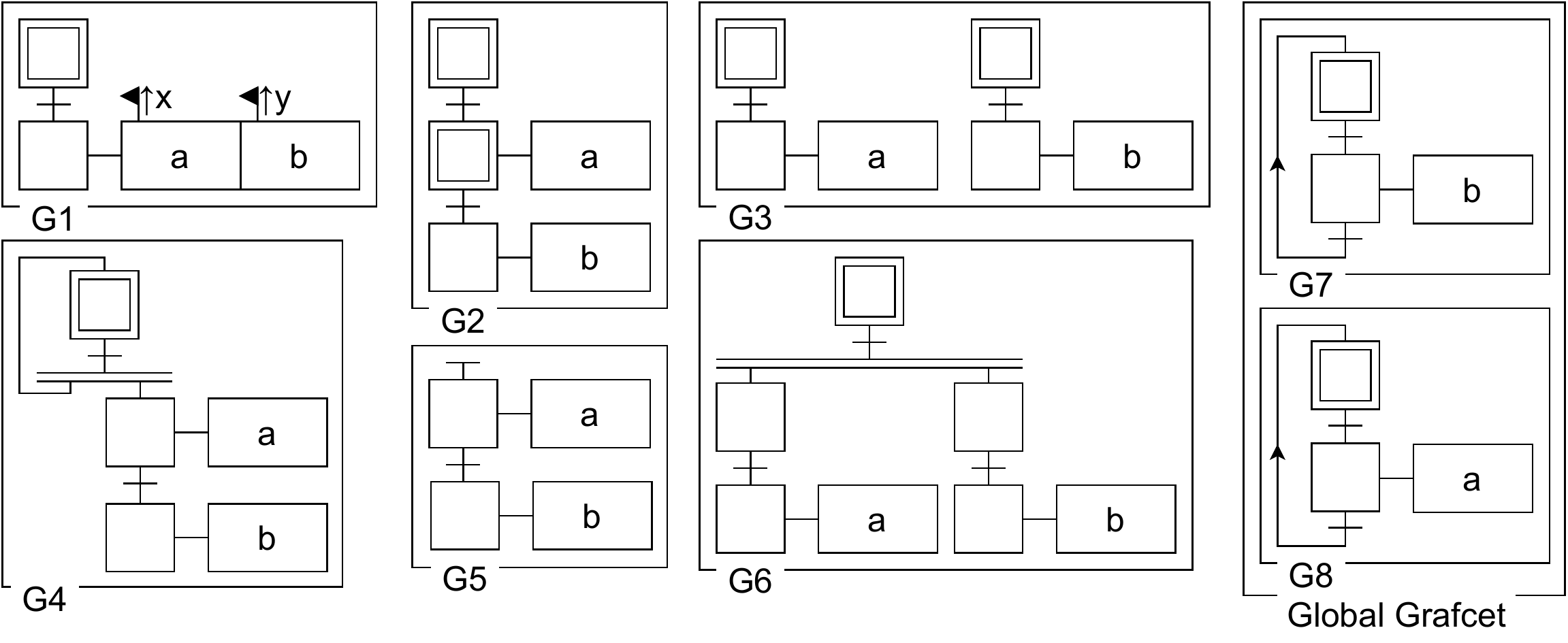}
	\caption{Different structures in GRAFCET \cite{iec60848} leading to concurrent behavior indicated by actions a and b.}
	\label{fig:concurrentStructures}
\end{figure*}

To apply abstract interpretation as shown in Section \ref{ssec:abstractInterpretation} to GRAFCET we first compare in Section \ref{ssec:problem} the control flow of GRAFCET to the control flow of sequential programs.
We point out that it is important to ensure the absence of concurrency in the Grafcet and, therefore, ensure soundness before applying a flow-sensitive abstract interpretation analysis presented in Section \ref{ssec:flowsens}.
As an abstract domain we choose intervals. 
With flow-sensitive we mean an analysis providing information about the variable values for every point in the control flow.

\subsection{Challenges of applying a control flow based analysis to GRAFCET}
\label{ssec:problem}
In GRAFCET different types of variables are defined \cite{iec60848}: 
Input variables are write-protected and are assumed to be non-deterministic since they correspond to sensors from the underlying process.
By applying abstract interpretation we approximate internal and output variables with elements from the abstract domain since internal variables have an influence on the state of the Grafcet and information about output variables can be useful to identify safety-critical situations.
Step variables can only be set by firings of transitions in accordance with the evolution rules of GRAFCET and therefore, are correlated to the control flow of the Grafcet. 
In the analogy to the CFG, the steps and transitions correlate to nodes in the CFG.

In a CFG built from a sequential program without concurrency, only one node of the CFG is executed at a time. 
This makes it easy to determine the execution order and execution number (e.g., how often an instruction is executed) of instructions on variables. 
However, the GRAFCET standard \cite{iec60848} presents different structures leading to concurrent behavior as shown in the partial Grafcets G1 to G8 in Fig. \ref{fig:concurrentStructures}:
\begin{itemize}
    \item Multiple conditional actions (graphically represented by a flag, followed by an expression like $\uparrow\! x$, where $\uparrow$ is called a rising edge of $x$ and occurs when $x$ changes from 0 to 1) associated to a single step (G1)
    \item Multiple initially active steps in sequence (G2) or parallel (G3)
    \item Elements producing active steps like source transitions ($\bullet t = \emptyset$ in G5) or its equivalence using an activation of parallel sequences (G4) as introduced by the standard \cite{iec60848}
    \item Activation of parallel sequences activating multiple steps at the same time ($|{t \bullet}| > 1$ in G6)
    \item Concurrently activated partial Grafcets (G7 and G8)   
\end{itemize}

All these structures can lead to a non-deterministic firing order of transitions and a non-deterministic execution order of actions. The latter is indicated in Fig. \ref{fig:concurrentStructures} by actions $a$ and $b$ in concurrent parts of the Grafcet. 
$a$ and $b$ here indicate any two types of actions that are dependent on each other. An example for $a$ could be $x := 0$, and an example for $b$ could be $x := x +1$, where the execution order has an influence on the resulting value of $x$.
Only the last structure G7 and G8 occurs in relation to a hierarchical structuring indicated by the Global Grafcet notation enclosing the partial Grafcets G7 and G8. 

All of the shown structures are not comparable to sequential control flow since statements are not executed concurrently in a single CFG. Multi-threading approaches that indeed deal with concurrency are usually based on multiple CFGs running concurrently to each other. However, every single CFG is sequential. Only partial Grafcets like in G7 and G8 are comparable to multi-threading programs.   

%%Häufigkeit
Besides the fact that the order of firing and execution is non-deterministic, their number of executions is non-deterministic as well. E.g., source transitions can non-deterministically generate multiple active steps in a sequence due to the non-deterministic change of input variables. Structures like shown in G4 in Fig.~\ref{fig:concurrentStructures} have a similar behavior.

In order to successfully apply a flow-sensitive abstract interpretation to GRAFCET, 
we first have to ensure the Grafcet being present does not behave concurrently with possible race conditions. 
%In other words, it has to be ensured none of the structures shown in Fig.~\ref{fig:concurrentStructures} are present.
The following conditions ensure for every possible partial Grafcet that it has no concurrent read and write instructions:
\begin{itemize}
    %G1
    \item     Each step has no associated conditioned actions $a \in A_\mathit{stor}$ with expressions $v_a$, $\mathit{val}_a$ and $b_a$ that depend on each other, corresponding to G1 in Fig.~\ref{fig:concurrentStructures}. 
    \item No multiple initially active steps are present ($|I_c| \leq 1 \land |M_c| \leq 1 \land |S_a| \leq 1$, the latter holds for all $a \in A_\mathit{fo}$), corresponding to G2 and G3 in Fig.~\ref{fig:concurrentStructures}. 
    \item No source transitions are present ($\bullet t \neq \emptyset$ holds for all $t \in T_c$), corresponding to G5 in Fig.~\ref{fig:concurrentStructures}.
    \item No variables are written in concurrency as a result of activation of parallel sequences as shown in G4 and G6 in Fig\ref{fig:concurrentStructures}  ($s_{a'} \notin S^C_{s_{a}} $ for all $v_{a'} = v_{a} \in V_{int} \cup V_{out}$ and $a', a \in A$, where $S^C_s \subseteq S$ is a set of steps concurrent to $s$. For the calculation of every $S^C_s$ with $s\in S$ we apply a structural analysis of the Grafcet)
\end{itemize}
Furthermore, to ensure the absence of concurrency, no internal and output variables must be written in multiple partial Grafcets concurrently ($ c = c'$ holds for all $ a \in A_c \cup A_{c'}$ and for all $v_a \in V_{int} \cup V_{out}$). 
If all the conditions presented above are met, a flow-sensitive analysis based on Alg.~\ref{alg:aiCFG} presented in Section \ref{ssec:flowsens} can be applied.

%========================================
%===========   Best Case   ==============  
%========================================

\subsection{Flow-sensitive abstract interpretation of GRAFCET}
\label{ssec:flowsens}
\begin{figure*}[t]
	\centering
	\includegraphics[width=.7\textwidth]{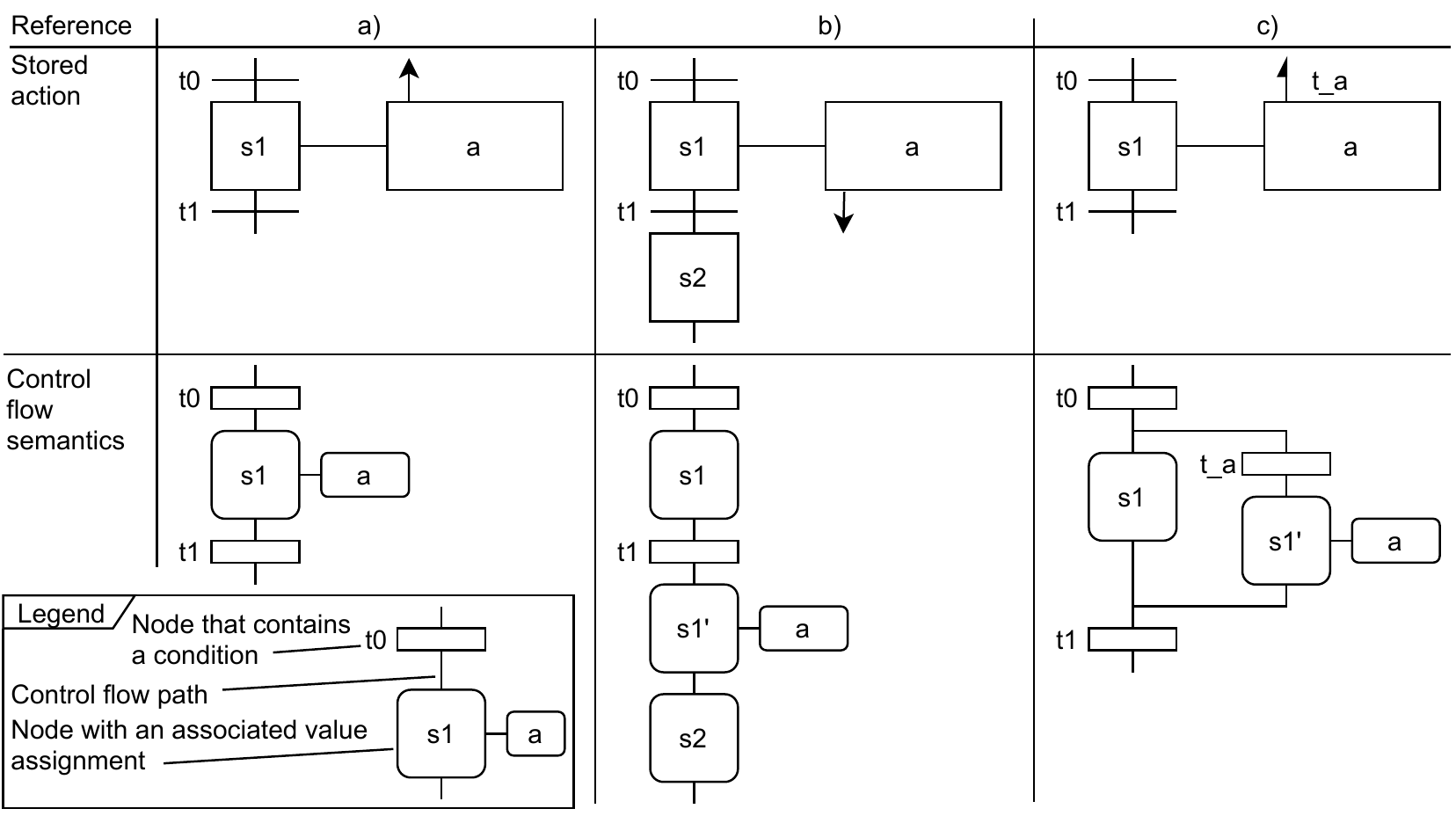}
	\caption{Control flow semantics of different stored action types.}
	\label{fig:normalized_actions}
\end{figure*}

Before applying the worklist algorithm for abstract interpretation over the control flow shown in Section \ref{ssec:abstractInterpretation} we need to define the control flow of GRAFCET.
In GRAFCET read instructions are connected to conditions associated with transitions (neglecting conditional actions for a moment) and write instructions are connected to actions associated with steps. Therefore, the statements of the control flow correspond to steps and transitions which are connected by arcs forming the flow relations.
By ensuring the absence of concurrency as described in Section \ref{ssec:problem}, we ensure the control flow to be sequential.
Although steps might not have an associated action and therefore do not change the state of the Grafcet the step's reachability could still be important information. %Streichpotenzial

To cover the different types of actions in GRAFCET, we need to consider them in more detail:
\begin{itemize}
    \item Continuous actions with a condition (might be true)
    \item Stored actions, activated by step activation (indicated by the upward arrow shown in Fig. \ref{fig:normalized_actions} a))
    \item Stored actions, activated by step deactivation (indicated by the downward arrow shown in Fig. \ref{fig:normalized_actions} b))
    \item Stored actions, activated by an event (indicated by the flag shown in Fig. \ref{fig:normalized_actions} c))
\end{itemize}
According to the standard \cite{iec60848}, the set of variables written in continuous actions and stored actions are disjoint %($\bigcup_{a \in A_\mathit{cont}} v_a \cap \bigcup_{a' \in A_\mathit{stor}} v_{a'} = \emptyset$) 
and therefore can be treated separately. 
Furthermore, continuous actions can only be applied on Boolean output variables.
The values of the corresponding Boolean output variables result implicitly from the corresponding step variables as well as the associated condition ($v = \bigvee_{a \in A_\mathit{cont}} x_{s_a} \land b_a$ with $v_a = v$). 
Therefore, they have no impact on the state of the Grafcet, so we ignore them for now. 
For stored actions, we apply a normalization, as shown in Fig. \ref{fig:normalized_actions}, before applying the abstract interpretation. Actions on activation are executed when the associated step is reached. Actions on deactivation are executed after the downstream transition is executed. For a single action activated by an event associated with a step, there are two possibilities. Either the condition might be true or false. This is modeled by an additional branch covering the condition of the action and the action itself. \\

\begin{algorithm}[H]
	\caption{Abstract interpretation of a sequential partial Grafcet}
	\label {alg:ai}
	\begin{algorithmic}[1]
  \Function{AbstInt}{$C$: partial Grafcet}
    \State $\mathit{Env}(n)$ is initialized to [0, 0] if $n \in I$, else to $\bot$
    \State $W \leftarrow I$
    \While {$W \neq \emptyset$}
        \State $n \leftarrow W.poll$
        \State $e \leftarrow \textsc{Transfer}(n, \mathit{Env}(n))$
        \ForAll{$n' \in \textsc{Succs}(C,n)$ such that $e \not\sqsubseteq \mathit{Env}(n')$}
            \State $\mathit{Env}(n') \leftarrow \mathit{Env}(n') \sqcup e$
            \State $W \triangleleft {n'}$
        \EndFor     
    \EndWhile
    \State \textbf{return} $\mathit{Env}$
    \EndFunction
	\end{algorithmic}
\end{algorithm}

We apply Alg. \ref{alg:aiCFG} for abstract interpretation on sequential Grafcets as shown in Alg. \ref{alg:ai}. $\mathit{Env}(n)$ is again the abstract environment (an abstract value for every variable in $V_\mathit{int} \cup V_\mathit{out}$) just before the read or write instructions associated with $n \in T \cup S$ are executed. To deal with only one variable type Boolean variables are modeled using integers with the usual interpretation that 0 corresponds to \textit{false} and 1 to \textit{true}.
According to the standard \cite{iec60848} integer variables are initialized to 0 and Boolean variables to \textit{false}. Therefore, we initialize the abstract values in $\mathit{Env}(n)$ for the initial step $n \in I$ to [0, 0] and for all other nodes in the Grafcet $n \in T \cup S$ to $\bot$. 
Input variables are not part of the abstract environment since they can change at all times to a non-deterministic value. Their abstract value therefore would be the trivial $\top$ element. The initial statement on the worklist is the initial step since $|I| = 1$. For the current  statement from the worklist we apply the function \textsc{Transfer}. The result of \textsc{Transfer} will be joined with the abstract environment $\mathit{Env}(n')$ of every successor statement $n'$ if it is not already included. If the calculated value of the abstract environment of the successor statement $n'$ is changing, it is put on the worklist.

%TRANSFER
Depending on whether $n \in S$ or $n \in T$ the function \textsc{Transfer} either executes the actions associated to the step in the abstract domain based on the approximated values in $Env(n)$ or applies the transition condition to $\mathit{Env}(n)$. 
For illustration, consider a step $n_i \in S$ with an associated action executing $Z := 0$ as soon as the step is activated. No matter what the approximation of $Z$ is at $n_i$, \textsc{Transfer}$(n_i, Env(n_i))$ returns [0, 0] as approximation for $Z$ after $n_i$ is executed. The approximation of all other variables does not change, since they are not affected by the value assignment. Further, consider a transition $n_{i+1} \in T$ with a condition $Z == 1$. \textsc{Transfer}$(n_{i+1}, Env(n_{i+1}))$ now returns $\bot$ as approximation for $Z$ after $n_{i+1}$ is executed, since there is no value in [0, 0] that satisfies the transition condition.

To deal with the Boolean operators (i.e. $\land, \lor, \neg$) in the abstract domain we substitute $X^\mathbb{B} \land Y^\mathbb{B}$ to $\alpha (X^\mathbb{Z}) \sqcap \alpha (Y^\mathbb{Z}), X^\mathbb{B} \lor Y^\mathbb{B} $ to $ \alpha (X^\mathbb{Z}) \sqcup \alpha (Y^\mathbb{Z}) $ and $ \neg X^\mathbb{B} $ to $ X^\mathbb{Z} == 0 $, where $\alpha(X)$ is the abstract value of the expression $X$ and 
$X^\mathbb{B}$ is the value of the Boolean expression in the Boolean domain which is substituted to $X^\mathbb{Z}$, the value in the integer domain.
Consider possible variable values $X^\mathbb{B} = Y^\mathbb{B} = \{\mathit{true}, \mathit{false}\}$ as an example at $n_j \in T$. In the abstract domain we get $\alpha(X^\mathbb{Z}) = \alpha(Y^\mathbb{Z}) = [0, 1]$ for $Env(n_j)$. The potential condition $X \land \neg Y$ of $n_j$ is substituted to $\alpha (X^\mathbb{Z} == 1) \sqcap \alpha (Y^\mathbb{Z} == 0)$.
Applying \textsc{Transfer}$(n_j, Env(n_j))$, the abstract value of the left hand side of the meet operator $\sqcap$ resolves to $\alpha(X^\mathbb{Z}) = [1, 1]$ and $\alpha(Y^\mathbb{Z}) = [0, 1]$ and the right hand side to $\alpha(X^\mathbb{Z}) = [0, 1]$ and $\alpha(Y^\mathbb{Z}) = [0, 0]$. Applying $\sqcap$ the return  value of \textsc{Transfer} is $\alpha(X^\mathbb{Z}) = [1, 1]$ and $\alpha(Y^\mathbb{Z}) = [0, 0]$,  which is the only possible approximation satisfying the transition condition.
%We assume without loss of generality that all expressions are in negation normal form.  

	\section{Evaluation}
\label{sec:eval}
The proposed approach was implemented and integrated in a toolchain developed by the authors. Part of the toolchain is a graphical editor for GRAFCET based on a GRAFCET meta-model proposed by Julius et al. \cite{Julius.19}. The meta-model was implemented using the Eclipse Modeling Framework (EMF)\footnote{\url{https://www.eclipse.org/modeling/emf/}}.  
For the abstract interpretation, we used the library Apron proposed by Jeannet et al. \cite{Jeannet.09}.

Besides the reachability of steps and transitions in combination with the respective approximation of internal and output variables, the analysis is able to detect problems regarding the firing of transitions: whether a transition can always fire or never fire which can lead to a deadlock or livelock, respectively, during run-time. Further it can detect so-called transient steps. A step is transient if an upstream and downstream transition evaluate to true in the same situation and therefore the transitions fire successively until a stable situation is reached when no transition can fire anymore.

The presented approach was evaluated using the GRAFCET-specification of an industrial plant first shown in \cite{Schumacher.14}. 
The application example is an automatic testing machine for quality control of components that consists of a conveyor belt, a rotary indexing table and six stations. Coordinated by the rotary indexing table, the parts pass through these stations, where separation and quality control take place. The components are marked as regular or damaged parts, and damaged parts are subsequently sorted out. 
The complete specification consists of 8 partial Grafcets, altogether consisting of 60 steps, 62 transitions, 46 stored actions, 15 continuous actions and 8 enclosing steps. In total 80 Boolean and integer variables are used (45 input, 20 output and 15 internal variables).
The duration of the analysis of the example applying Alg.~\ref{alg:ai} was between 20 and 500 milliseconds per partial Grafcet and about 1200 milliseconds in total. By applying the analysis, we were able to identify some unreachable steps in the specification, which was due to an input error in one of the transitions.
We compared the duration of the control flow based approach proposed in this work to the model checking approach proposed in \cite{Mross.22}. Using the model checking approach \cite{Mross.22} we checked every partial Grafcet on its own for reachability of the steps using a CTL formula according to the scheme $EF(x_s)\land EF(x_{s+1}) \land \dotsb$. 
The duration of the analysis using the model checking approach was between 7 milliseconds and 129 seconds per partial Grafcet and about 135 seconds in total. This test confirms that the presented control flow based approach scales more effectively for industrial sized GRAFCET specifications.

\begin{figure}[t]
	\centering
	\includegraphics[width = \columnwidth]{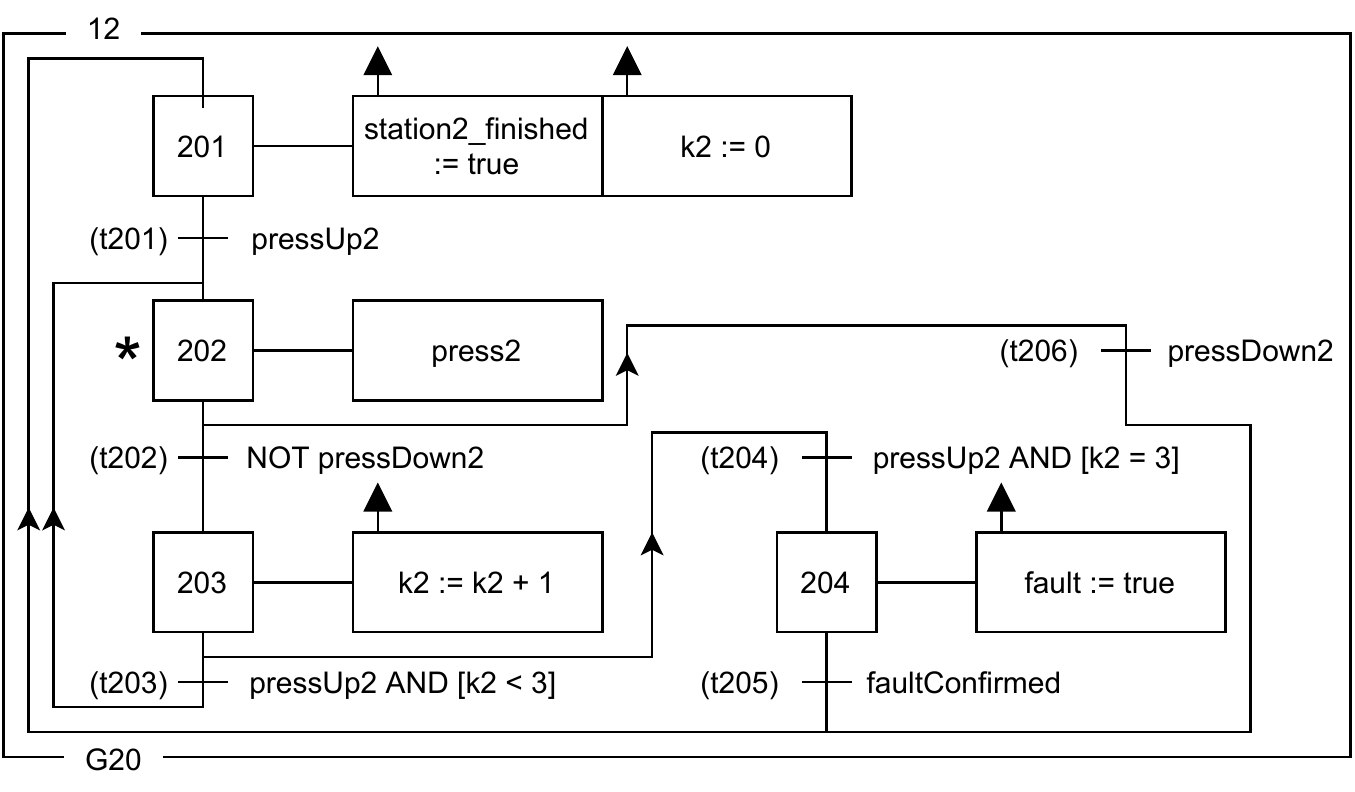}
	\caption{Application example Grafcet G20.}
	\label{fig:g2}
\end{figure}

In the remainder of the section we illustrate the results on the GRAFCET specification G20 of the second station of the illustration example, shown in Fig.~\ref{fig:g2}\footnote{The full specification formalized with GRAFCET can be viewed here: \url{https://github.com/Project-AGRAFE/GRAFCET-instances}}
An emergency stop as well as the coordination of the stations is implemented using an enclosing steps that can start and stop the partial Grafcet G20. In Fig. \ref{fig:g2} the number 12 at the top refers to the enclosing step 12 controlling the station. The asterisk at step 202 marks that the step is activated by the superior enclosing step. The station has the task to fix the parts into the socket of the rotary indexing table. The fix is done by a piston accessed by the variable \textit{press2}. The part is pressed into the socket up to three times before the station reports a fault. Otherwise the station reports the termination of the process setting \textit{station2\_finished} to true. The counting of the attempts is done by using the internal variable \textit{k2}. 
\begin{table}[t]
    \centering
    \caption{Results of the control flow based analysis of G20 shown in Fig. \ref{fig:g2}.}
    \label{tab:resultG2}
    \begin{tabular}{c  c  c  c }
    Step/Transition & $k2$ & $station2\_finished$ & $fault$\\\hline
Step 201 &		[0,	3] & [0, 1]& [0, 1]\\
Step 202 &		[0,	2] & [0, 1]& [0, 1]\\
Step 203 &		[0, 2] & [0, 1]& [0, 1]\\
Step 204 &		[3, 3] & [0, 1]& [0, 1]\\

Transition t201 &	[0,	0] & [1, 1]& [0, 1]\\
Transition t202 &	[0,	2] & [0, 1]& [0, 1]\\
Transition t203 &	[1,	3] & [0, 1]& [0, 1]\\
Transition t204 &	[1,	3] & [0, 1]& [0, 1]\\
Transition t205 &	[3,	3] & [0, 1]& [1, 1]\\
Transition t206 &	[0,	2] & [0, 1]& [0, 1]
    \end{tabular}
\end{table}

The analysis starts by initializing the internal and output variables to 0 for step 202 and to $\bot$ for all other steps and transitions. The algorithm iterates over the preceding steps and transitions in G20, resulting in the approximated intervals for the variables \textit{k2}, \textit{station2\_finished} and \textit{fault}, shown in Table \ref{tab:resultG2}. The results show that every step and transition is reachable since the corresponding abstract values differ from the initialized value $\bot$, which is the expected behavior.
	\section{Conclusion}
\label{sec:conclusion}
The goal of this paper was to present an automatic and scalable analysis approach for IEC 60848 GRAFCET compared to other approaches like model checking. Therefore, we chose a static analysis approach based on the control flow of GRAFCET. 
In Section \ref{ssec:problem} we compared the control flow of GRAFCET to a sequential CFG in order to adapt analysis means applicable to a CFG. We pointed out why this is possible only for Grafcets without concurrent behavior. In order to apply a sequential analysis we presented what elements of GRAFCET result in concurrent behavior and how to identify these to ensure the absence of concurrency. 

The proposed analysis itself approximates the variable values for every step and transition in the control flow using abstract interpretation.
This results in an analysis result that has the same structure like the Grafcet itself, making the remediation of design errors easier in comparison to inspecting a complete state space. 
On a realistic example the evaluation has revealed that run-time errors like, e.g., unreachable states can be detected. 
Further, the evaluation has revealed that Grafcets of realistic size can be analyzed in a reasonable amount of time.
In comparison to the model checking approach presented in \cite{Mross.22} the scalability of the analysis is improved. However, with the disadvantage that an approximation in general can lead to false alarms and the properties to be checked are limited to safety properties.

Since the presented approach is applicable only for a subset of GRAFCET without concurrency we currently work on extending the approach to deal with concurrent partial Grafcets. This could be achieved by adapting algorithms proposed for multi-threaded programs, such as those by Kusano et al. \cite{Kusano.16}. In addition, other abstract domains could be used besides the interval lattice to achieve a more precise approximation of the variable values.
	
	\begin{acknowledgments}
		This research is part of the project "Analysis of GRAFCET specifications to detect design flaws" (project number 445866207) funded by the Deutsche Forschungsgemeinschaft.
	\end{acknowledgments}
	
	\bibliographystyle{IEEEtran.bst}
	\bibliography{bibliography}
	
\end{document}